\RequirePackage{ifpdf}
\ifpdf 
\documentclass[pdftex]{sigma}
\else
\documentclass{sigma}
\fi

\numberwithin{equation}{section}

\begin{document}

\allowdisplaybreaks

\renewcommand{\thefootnote}{$\star$}

\renewcommand{\PaperNumber}{031}

\FirstPageHeading

\ShortArticleName{A Recurrence Relation Approach to  Higher Order Quantum Superintegrability}

\ArticleName{A Recurrence Relation Approach\\ to  Higher Order Quantum Superintegrability\footnote{This paper is a
contribution to the Special Issue ``Symmetry, Separation, Super-integrability and Special Functions~(S$^4$)''. The
full collection is available at
\href{http://www.emis.de/journals/SIGMA/S4.html}{http://www.emis.de/journals/SIGMA/S4.html}}}

\Author{Ernie G.~KALNINS~$^\dag$,  Jonathan M.~KRESS~$^\ddag$  and Willard  MILLER  Jr.~$^\S$}

\AuthorNameForHeading{E.G.~Kalnins, J.M.~Kress  and  W.~Miller Jr.}

\Address{$^\dag$~Department of Mathematics,  University of Waikato, Hamilton, New Zealand}
\EmailD{\href{mailto:math0236@math.waikato.ac.nz}{math0236@math.waikato.ac.nz}}
\URLaddressD{\url{http://www.math.waikato.ac.nz}}
\Address{$^\ddag$~School of Mathematics, The University of New South Wales,
Sydney NSW 2052, Australia}
\EmailD{\href{mailto:j.kress@unsw.edu.au}{j.kress@unsw.edu.au}}
\Address{$^\S$~School of Mathematics, University of Minnesota,
 Minneapolis, Minnesota, 55455, USA}
\EmailD{\href{mailto:miller@ima.umn.edu}{miller@ima.umn.edu}}
\URLaddressD{\url{http://www.ima.umn.edu/~miller/}}

\ArticleDates{Received January 27, 2011, in f\/inal form March 20, 2011;  Published online March 28, 2011}

\Abstract{We   develop  our method to prove quantum
superintegrability of an integrable  2D system, based on
recurrence relations obeyed by
the eigenfunctions of the system with respect to separable coordinates. We
show that the method provides rigorous proofs of superintegrability and explicit
constructions of  higher order  generators for the symmetry algebra.
We apply the method to 5 families of systems, each depending on a para\-me\-ter~$k$, including
most notably the caged anisotropic oscillator, the  Tremblay,
Turbiner and Winternitz system and a deformed Kepler--Coulomb system, and we give
proofs of quantum superintegrability for all
rational values of $k$, new for 4 of these systems. In addition, we show
that the explicit  information supplied by the special function recurrence
relations  allows us to prove, for the f\/irst time in 4 cases, that the symmetry algebra generated by our lowest order
symmetries
closes and  to determine the associated  structure equations of the algebras for each $k$. We have no proof  that our generating symmetries are of lowest possible order, but we have no counterexamples, and we are conf\/ident we can can always f\/ind any missing generators from our raising and lowering operator recurrences.  We
also get for free, one variable models of the
action of the symmetry algebra in terms of dif\/ference operators. We describe how the St\"ackel transform acts and show
that it preserves the structure equations.}


\Classification{20C99; 20C35; 22E70}

\section{Introduction}

An $n$-dimensional quantum (maximally) superintegrable system is an integrable
Hamiltonian
system that not only possesses $n$ mutually  -- commuting symmetry operators, but
in addition,
the Hamiltonian commutes with  $n-1$ other f\/inite order partial dif\/ferential
operators such that the
 $2n-1$ operators are algebraically independent.
We restrict to quantum systems of the form $H=\Delta_n+ V$ where
$\Delta_n$ is the Laplace--Beltrami operator on some real
 or complex
Riemannian manifold and $V$ is a potential function locally def\/ined on the manifold. These
systems, inclu\-ding the
 quantum anisotropic oscillator and the hydrogen atom have great historical
importance, due to
their remarkable properties, \cite{SCQS,IMA}. The order of a quantum
superintegrable system is the
maximum order  of the quantum symmetries as dif\/ferential operators (with the
Hamiltonian excluded). However,
we always choose the generators such that the maximum order is as small as
possible.  Systems of 2nd order
have been well studied and there is now a structure and classif\/ication theory
\cite{ KKM20061,KKMP, DASK2005, KKM2007,KMP2007,KMP2008}.
Until very recently, there were comparatively few known superintegrable systems
of order $\ge3$ and virtually no
structure theory for the algebra generated by the symmetries. In the last three
years, however, there has been a dramatic increase
in discovery of new families of possible higher order superintegrable classical
and quantum systems
\cite{CDR2008, RTW, Evans2008a,TTW,TTW2,CQ10,TW2010, PW2010}.
The authors and collaborators, and others,  have developed methods for verifying
superintegrability of these proposed systems,
\cite{BH, KMP10,KKM10,KKM10a,KMPog10}. The f\/irst method developed by us, to verify
superintegrability
for  2-dimensional  quantum systems, was based on a~canonical form
for symmetry operators of arbitrary order. This approach succeeded for several
important systems, such as the caged
anisotropic oscillator and the TTW potential for all rational $k=p/q$, but it
led to multi-term recurrence formulas for which very careful analysis
was needed to verify f\/inite dimensional solution spaces, \cite{KKM10a, KKM10b}. Furthermore,
the approach yielded no information about the
structure of the  algebras generated by the symmetries. In  \cite{KKM10b} the
authors introduced a new method, based on
the recurrence relations obeyed by the separated eigenfunctions of the
Hamiltonian, and sketched its application to the caged
anisotropic oscillator. In this paper we further develop and make rigorous the
special function recurrence relation method,
and apply it  to additional cases, including the TTW system. We show that this
new method enables us to prove for the f\/irst time
that the symmetry algebra for  the TTW
potential is closed for all rational $k$, and to compute
the structure relations for the algebra. (In a series of recent papers \cite{Marquette20101,Marquette20102, Marquette20091,
Marquette20103}  Marquette has used a ladder operator method to determine higher order symmetry operators and structure equations
for 2D superintegrable systems that separate in Cartesian coordinates. In particular, he has found the structure equations for the caged anharmonic
oscillator. Our method is related to his for separation in Cartesian coordinates, but more general
 in that it applies to systems that separate in any
coordinate system, e.g., polar coordinates.)

The basic idea that motivates the method is that, since the formal eigenspaces of
the Hamiltonian are invariant under action of
any  symmetry operator, the operator must induce recurrence relations for the
basis of  eigenfunctions corresponding to the associated
coordinate system in which the eigenvalue equation separates. The recurrence
relation method uses the known recurrence relations
for hypergeometric functions in order to reverse this process and determine a symmetry
operator from a suitable combination of
 recurrence relations. We can compute the symmetry operators and structure
equations for the symmetry algebra by restricting ourselves
to a formal ``basis'' of separated eigenfunctions. Then we appeal to our theory
of  canonical forms for symmetry
operators to show that results
obtained on restriction to a formal eigenbasis actually hold as true identities
for purely dif\/ferential operators
def\/ined independent of ``basis'' functions. (We have no proof that this reverse process will always work but conjecture that it will succeed whenever all the separated eigenfunctions are of hypergeometric type.)  We start with a simple example on
the 2-sphere, to introduce the theory.
Then we consider an example in
Minkowski space followed by a revisiting of the caged anisotropic oscillator.
Finally we treat the TTW potential again.
In four of  these cases we give the f\/irst proofs of the closure and the structure of
the symmetry algebras.

\section{A simple system} \label{system1}

Here we construct a  proof of quantum integrability using recurrence formula
techniques. As a~trial system we consider a  quantum Hamiltonian  on the
two-sphere:
\begin{gather} \label{ham1} H=\partial ^2_\theta +\cot\theta \partial _\theta + \frac{1}{
\sin ^2\theta }
\partial ^2_\varphi  + \frac{\alpha }{ \sin ^2\theta \cos^2k\varphi },
\end{gather}
where $k=\frac{p}{ q}$ and $p$, $q$ are relatively prime positive integers. (Here, all our considerations are local. We do not require that the potential is globally  def\/ined  on the two-sphere, or that there are any boundary conditions. Global issues can be examined on a case-by-case basis. Also, all parameters and variables can be complex, except for~$k$ which is rational.) This system is clearly integrable, since it
admits a 2nd order symmetry, responsible
for separation of the eigenvalue equation for $H$ in spherical coordinates. Our aim
is to show that
this Hamiltonian admits an additional quantum symmetry, so is superintegrable. This is relatively
straightforward to do using recurrence formula techniques. The typical
separable solution in spherical coordinates has the form
\[
\Psi =T^{k(N+\frac{1}{ 2})}_n(\cos\theta )(\cos\psi )^{1/2} U^a_N(\sin\psi ),
\]
where $T,U=P$ or $Q$ are the solutions of Legendre's equation,
$\alpha =k^2(\frac{1}{ 4}-a^2)$ and $\psi =k\varphi$. Indeed if we write $\Psi =
\Theta (\theta )\Phi (\varphi )$ then these  solutions satisfy the
``eigenfunction'' equations  $H\Psi=-n(n+1)\Psi$ where $H=L_1$ and
$L_2\Psi=-k^2(N+\frac12)^2\Psi$, or
\[
\left(\partial ^2_\theta +\cot\theta \partial _\theta -
\frac{k^2(N+\frac{1}{ 2})^2}{ \sin ^2\theta }\right)\Theta  = -n(n+1)\Theta \]
and
\[\left(\partial ^2_\varphi + \frac{k^2(\frac{1}{ 4}-a^2)}{ \cos ^2k\varphi } \right)\Phi
=
-k^2\left(N+\frac{1}{ 2}\right)^2\Phi,
\]
where
\[\Psi=\Theta (\theta )\Phi (\varphi )=T^{k(N+\frac{1}{ 2})}_n(\cos\theta
)(\cos\psi )^{
1/2} U^a_N(\sin\psi ).\]
 (Note that $L_2$ is the symmetry operator  associated with
variable separation in spherical coordinates.) Here no boundary conditions are
implied and all parameters can be complex.
To proceed further we observe the following recurrence
formulas for Legendre functions~\cite{AAR}:
\begin{gather*}
D^+_\nu (x)T^\mu _\nu (x)=\big(1-x^2\big)\frac{\partial }{
\partial x}
T^\mu _\nu (x)-(\nu +1)xT^\mu _\nu (x)=-(\nu -\mu +1)T^\mu _{\nu +1}(x),\\
  D^-_\nu (x)T^\mu _\nu (x)=\big(1-x^2\big)\frac{\partial }{ \partial x}
T^\mu _\nu (x)+\nu xT^\mu _\nu (x)=(\nu +\mu )T^\mu _{\nu -1}(x)\nonumber
\end{gather*}
and
\begin{gather*}
C^+_\mu (x)T^\mu _\nu (x)=\big(1-x^2\big)^{1/2}\frac{\partial }{ \partial x}
T^\mu _\nu (x)+ \frac{\mu x}{ (1-x^2)^{1/2} }T^\mu _\nu (x)=-T^{\mu +1}_\nu
(x),\\
  C^-_\mu (x)T^\mu _\nu (x)=\big(1-x^2\big)^{1/2}\frac{\partial }{ \partial x}
T^\mu _\nu (x)- \frac{\mu x}{ (1-x^2)^{1/2} }
T^\mu _\nu (x)=(\nu+\mu )(\nu -\mu +1)T^{\mu -1}_\nu (x).\nonumber
\end{gather*}
These relations enable the shifting of the indices $\nu $ and $\mu \ by
\pm 1$. There are clearly similar relations for the functions
$U^\mu _\nu (y)$. If we  let  $N\rightarrow N+q$
then
\[\Psi  \rightarrow  T^{k(N+\frac{1}{ 2})+p}_n(\cos\theta )(\cos\psi )^{1/2}
U^a_{N+q}(\sin\psi ).\]

We note that $N$ can be quite arbitrary and now consider the function
\[\Upsilon _+= T^{k(N+\frac{1}{ 2})+p}_n(x) U^a_{N+q}(y),
\]
where $x=\cos\theta $ and $y=\sin\psi$. This function can be obtained from
\[\Upsilon _0= T^{k(N+\frac{1}{ 2})}_n(x) U^a_N(y)\]
via raising operators:
\begin{gather} \Delta _+(N)\Upsilon _0=C^+_{k(N+\frac12)+p-1}(x)\cdots
C^+_{k(N+\frac12)}
(x)D^+_{N+q-1}(y)\cdots D^+_N(y)\Upsilon _0\nonumber\\
\phantom{\Delta _+(N)\Upsilon _0}{}  =(-1)^{p+q}\left(N-a+1\right)_q\Upsilon _+.\label{raising1}
 \end{gather}
Here $(\alpha)_\ell=\alpha(\alpha+1)\cdots(\alpha+\ell-1)$ is the Pochhammer
symbol, and we assume that we have chosen bases for
 the separable solutions such that the same recurrence formulas and
normalizations
hold for {\it all} elements of the basis. Thus relation~(\ref{raising1}) holds
for all four elements in the basis.
Similarly, we look at the possibility that $N \rightarrow  N-q$. This can be
obtained
from $\Upsilon _0$ via lowering operators:
\begin{gather}
\Delta _-(N)\Upsilon _0=C^-_{k(N+\frac12)-p+1}
(x)\cdots C^-_{k(N+\frac12)}(x)D^-_{N-q+1}(y)\cdots D^-_N(y)\Upsilon _0\nonumber\\
 \phantom{\Delta _-(N)\Upsilon _0}{} =
(-1)^{p+q}\left(-N-a\right)_q\left(-n-k\left(N+\frac12\right)\right)_p\left(n-k\left(N+\frac12\right)
+1\right)_p\Upsilon _-,\label{lowering1}
\end{gather}
where
\[\Upsilon _-= T^{k(N+\frac{1}{ 2})-p}_n(x) U^a_{N-q}(y).\]
This result is independent of which basis we choose.

Now consider the dif\/ferential operator $\Delta =\Delta _+ +\Delta _-$. It
follows from the form of the $C$
and $D$ operators that this is an even  polynomial function of $N+\frac{1}{ 2}$.
This can be seen from the $\nu
\rightarrow  -\nu -1$ symmetry  of the operators $D$ and the $\mu
\rightarrow  -\mu $ symmetry of the operators $C$: $C^+_\mu (x) = C^-_{-\mu
}(x)$, $D^+_\nu (y) = D^-_{-\nu -1}(y)$. Note that
 $(N+\frac{1}{ 2})^2\Psi =-k^{-2}\big(\partial ^2_\varphi +
\frac{k^2(\frac{1}{ 4}-a^2)}{ \sin ^2k\varphi } \big)\Psi$ for any eigenfunction $\Psi$ of $L_2$. It follows that under the
transformation $N\rightarrow -N-1$
we have $\Delta _+\rightarrow \Delta _-$ and $\Delta _-\rightarrow \Delta _+$.
Thus, everywhere the term $(N+\frac12)^{2\ell}$
occurs in the expansion of $\Delta$ (where $\ell$ is a positive integer) we can
replace it by $(k^{-2}L_2)^\ell$
and obtain a pure dif\/ferential operator, independent of the parameters~$n$,~$N$. As
a consequence we see that
we have constructed a pure dif\/ferential operator, which we also call $\Delta$,
and  which preserves each eigenspace of
$H$ when acting on functions of the form
\[\Upsilon _0= T^{k(N+\frac{1}{ 2})}_n(x) U^a_N(y)=\Psi (\sin\psi )^{-1/2}.\]
Similarly, the operator $\Delta_+-\Delta_-$ goes to its negative under the
mapping $N\rightarrow -N-1$, so it is
an odd function of $N+\frac12$. This implies that ${\tilde
\Delta}=(1/(N+\frac12))(\Delta_+-\Delta_-)$
is an even function of $N+\frac12$, so it can also be represented as a
pure dif\/ferential operator, independent of the parameters~$N$,~$n$ and it preserves
each eigenspace of~$H$.
We have constructed two partial dif\/ferential opera\-tors~$\Delta$ and~$\tilde
\Delta$, each of which commutes with the
Hamiltonian $H$ on formal eigenspaces. Thus they act like symmetry operators.
However, to prove this we must show that they commute
with $H$ when acting on any analytic functions, not just eigenfunctions. To
establish this fact we use the canonical form for
symmetry operators studied in our papers \cite{KKM10a, KKM10b}.

Although we give the reasoning for this special system, our argument is quite
general and immediately applicable to other systems.
The operator~$L_2$ determines  separable coordina\-tes~$x$,~$y$ for the system (in
this case $(x,\psi)$.
Now consider the commutator $[H,\Delta]$. When acting on formal eigenfunctions
$\Upsilon _0$ the commutator gives~0.
We want to show that it vanishes identically. To do this we write  $[H,\Delta]$
in canonical form by recursively replacing all
second derivatives~$\partial^2_x$,~$\partial^2_y$ in terms of $H$, $L_2$ to obtain
the expression
\[[H,\Delta ] =A(x,y,H,L_2)\partial^2 _{xy}+B(x,y,H,L_2)\partial
_x+C(x,y,H,L_2)\partial _y+D(x,y,H,L_2).\]
This expression has to be interpreted as in \cite{KKM10a}, i.e., the parameters
$H$, $L_2$ must be moved to the right before being
identif\/ied as operators.
Now applying this operator to any eigenfunction we obtain
\begin{gather}\label{4eqns4unkowns}[H,\Delta ]\Upsilon _0 =\left(A\partial^2
_{xy}+B\partial _x+C\partial _y+D\right)\Upsilon _0=0,\end{gather}
for all choices of the parameters $H$, $L_2$. Noting that we have  4 linearly
independent choices for $\Upsilon _0=P_i(x)Q_j(y)$,
 we can write
(\ref{4eqns4unkowns}) as a set of 4 homogeneous equations for the 4 unknowns
$A$, $B$, $C$, $D$:
\[ \left(  \begin{array}{llll} P'_1(x)Q'_1(y) & P'_1(x)Q_1(y) & P_1(x)Q'_1(y) &
P_1(x)Q_1(y)\\
P'_1(x)Q'_2(y) & P'_1(x)Q_2(y) & P_1(x)Q'_2(y) & P_1(x)Q_2(y)\\
 P'_2(x)Q'_1(y) & P'_2(x)Q_1(y) & P_2(x)Q'_1(y) & P_2(x)Q_1(y)\\
 P'_2(x)Q'_2(y) & P'_2(x)Q_2(y) & P_2(x)Q'_2(y) & P_2(x)Q_2(y)\end{array}\right)
\left( \begin{array}{l} A\\B\\C\\D\end{array}\right)=\left(\begin{array}{l}0\\0\\0\\0\end{array}\right).
\]
It is convenient to introduce the determinant
function
\begin{gather*}
W(f_1(x),g_1(y),f_2(x),g_2(y))=
  \left| \begin{array}{llll} f'_1(x)g'_1(y) & f'_1(x)g_1(y) & f_1(x)g'_1(y) &
f_1(x)g_1(y)\\
f'_1(x)g'_2(y) & f'_1(x)g_2(y) & f_1(x)g'_2(y) & f_1(x)g_2(y)\\
 f'_2(x)g'_1(y) & f'_2(x)g_1(y) & f_2(x)g'_1(y) & f_2(x)g_1(y)\\
 f'_2(x)g'_2(y) & f'_2(x)g_2(y) & f_2(x)g'_2(y) & f_2(x)g_2(y)\end{array}\right| \nonumber\\
\phantom{W(f_1(x),g_1(y),f_2(x),g_2(y))}{}
 =\left|\begin{array}{ll} f_1(x)& f_2(x)\\ f_1'(x)& f_2'(x)\end{array}\right|^2\cdot
\left|\begin{array}{ll} g_1(y)& g_2(y)\\ g_1'(y)&g_2'(y)\end{array}\right|^2.
\end{gather*}
We then note that  for our particular system
\[W(P_1(x),P_2(x),Q_1(y),Q_2(y))\neq 0,\]
except at isolated points, since the Wronskian of two independent solutions of a
separated eigenfunction equation is nonzero.
Thus we conclude that  $A=B=C=D=0$. Consequently  $[H,\Delta ]=0$ identically.
This proves that $\Delta=L_3$ is a symmetry operator for the system. The same
proof shows that $\tilde \Delta=L_4$ is also a
symmetry operator. Since both $L_3$ and $L_4$ fail to commute with $L_2$,
each must be algebraically independent of $H=L_1,L_2$. Thus this system is
superintegrable for all rational~$k$.

\begin{example} \label{example1} If $k=p=q=1$ we have the familiar superintegrable system on
the 2-sphere where the corresponding Schr\"odinger operator is
\[H =\partial ^2_\theta +\cot\theta \partial _\theta +
\frac{1}{ \sin ^2\theta } \partial ^2_\varphi  +
\frac{\alpha }{ \sin ^2\theta \cos^2\varphi}, \]
i.e., the Laplacian on the two sphere $s^2_1+s^2_2+s^2_3=1$ (in Cartesian
coordinates $s_j$)
plus the potential $V({\bf s})=\frac{\alpha }{ s^2_1}$. This corresponds to system [S3] in the list \cite{KKMP}.
Then we have
$L_3=\Delta _++\Delta _-$ and $L_4= \frac{1}{ (N+\frac{1}{ 2})}
(\Delta _+-\Delta _-)$ with $\Delta _+$ and $\Delta _-$ where
\begin{gather*}
\Delta _+(N)f=\big(\big(1-y^2\big)\partial _y-(N+1)y\big)\left(\sqrt{ 1-x^2}\partial _x+
\frac{(N+\frac{1}{ 2})x}{ \sqrt{1-x^2}}\right)f
\\
\phantom{\Delta _+(N)f}{}   =(\cos\psi \partial _\psi -(N+1)\sin\psi )\left(-\partial _\theta +\left(N+\frac{1}{ 2}\right)
\cot\theta \right)f,\\
\Delta _-(N)f=\big(\big(1-y^2\big)\partial _y+Ny\big)\left(\sqrt {1-x^2}\partial _x-
\frac{(N+\frac{1}{ 2})x}{ \sqrt{1-x^2}}\right)f\\
\phantom{\Delta _-(N)f}{}
=(\cos\psi \partial _\psi +N\sin\psi )\left(-\partial _\theta -\left(N+\frac{1}{ 2}\right)
\cot\theta \right)f,
\end{gather*}
where $x=\cos\theta $ and $y=\sin\varphi$. Note that in this example
$\psi =\varphi $ since
 $k=1$.

We choose the standard spherical coordinates on the sphere viz.
\[s_1=\sin\theta \cos\varphi  ,\qquad s_2=\sin\theta \sin\varphi  ,\qquad s_3=\cos\theta \]
and the corresponding expressions
\[
J_1=s_2\partial _{s_3}-s_3\partial _{s_2}, \qquad
J_2=s_3\partial _{s_1}-s_1\partial _{s_3},\qquad
J_3=s_1\partial _{s_2}-s_2\partial _{s_1}.
\]
The Hamiltonian can then be written as
\[H=J^2_1+J^2_2+J^2_3+\frac{\alpha }{ s^2_1}.\]

From these calculations we deduce that
\[
L_4=-2J_1,\qquad L_3=\{J_2,J_3\}+J_1+2\alpha  \frac{s_2s_3}{ s^2_1},
\]
where the separation equation in polar coordinates is due to
$L_2=J^2_3+ {\alpha }/{ \cos ^2\psi }$
with eigenvalue $-(N+\frac{1}{ 2})^2$. (Here, $\{A,B\}=AB+BA$.)
We also note the relation  $2L_3+L_4=[L_4,L_2]$.
Thus we see that $[L_4,L_2]$ is not proportional to $L_3$.
\end{example}

We can compute the general structure relations for the symmetries of system~(\ref{ham1}). We obtain
\begin{gather*}
[L_2,L_4]=-2qk^2L_3-q^2k^2L_4,\\
 [L_2,L_3]= -q^2k^2L_3+2qL_4L_2=q^2k^2L_3+q^3k^2L_4+q\{L_2,L_4\}.
\end{gather*}

To compute $[L_3,L_4]$ we need some preliminary results.  We make note of the
identities
\begin{gather*}
\Delta _+ (N-q)\Delta _-(N)\Upsilon _0=(-1)^q(a-N)_q(-N-a)_q\\
\qquad{} \times \left(-n-k\left(N+\frac{1}{
2}\right)\right)_p\left(n-k\left(N+\frac{1}{ 2}\right)+1\right)_p\Upsilon _0=F_1(n,N)\Upsilon _0,\\
 \Delta _-(N+q)\Delta _+(N)\Upsilon _0
=(-1)^q\left(N-a+1\right)_q\left(N+a+1\right)_q\\
\qquad {} \times \left(-n+k\left(N+\frac{1}{
2}\right)\right)_p \left(n+k\left(N+\frac{1}{ 2}\right) +1\right)_p\Upsilon _0=F_2(n,N)\Upsilon _0.
\end{gather*}
Recall that $E=-n(n+1)$ is the eigenvalue of the Hamiltonian and $-k^2(N+\frac12)^2$ is the eigenvalue of $L_2$ corresponding to a basis function. Using the property
$(-\alpha)_\ell=(-1)^\ell(\alpha-\ell+1)_\ell$
we note that $F_1(n,-N-1)=F_2(n,N)$ and $F_j(-n-1,N)=F_j(n,N)$ for $j=1,2$.
Consequently,
$F^+=F_1(N,n)+F_2(N,n)$
 is an even polynomial function in $N+\frac{1}{ 2}$ and a polynomial function of
$n(n+1)$, hence when acting on
separated basis functions  $F^+=P^+(H,L_2)$ is
 a polynomial in $H$ and $L_2$. In fact, $F^+=P(H,L_2)$ as pure dif\/ferential
operators,
independent of basis. The proof of this fact is analogous to that given above.
We write the operator $F^+ -P^+$ in canonical form
$A\partial^2_{xy}+b\partial_x+C\partial_y+D$. Then on an arbitrary eigenbasis we have
\[
\left(A\partial^2 _{xy}+B\partial _x+C\partial _y+D\right)\Psi_n=0.
\]
It follows via the usual Wronskian argument that $A=B=C=D=0$. Thus $F^+-P^+=0$
identically. Similarly
$F^-=\frac{1}{N+\frac12}(F_1(n,N)-F_2(n,N))$ is an even polynomial function in
$N+\frac{1}{ 2}$ and a
polynomial function of $n(n+1)$, so   $F^-=P^-(H,L_2)$ is
 a polynomial in~$H$ and~$L_2$.

Now it is straightforward to obtain
\[
[L_3,L_4]=qL_4^2-2P^-(H,L_2).
\]
 There is of course an extra constraint. In
fact
\[ L_4^2L_2=-k^2L_3^2+qk^2L_4L_3+2k^2P^+(H,L_2),\]
and, symmetrizing, we f\/ind
\[\{L_4,L_4,L_2\}=-6k^2L_3^2-q^2k^2L_4^2-3qk^2\{L_3,L_4\}-10qk^2P^-(H,
L_2)+12k^2P^+(H,L_2),
\]
where $\{A,B,C\}$ is the symmetrizer of 3 operators.
The structure is clearer if we def\/ine $R=-2qk^2L_3-q^2k^2L_4$. Then we can
rewrite the structure equations as
\begin{gather*}
 [L_2,L_4]=R,\\
 [L_2,R]=-2q^2k^2\{L_2,L_4\}-q^4k^4L_4,\\
 [L_4,R]=2q^2k^2L_4^2-4qk^2P^-(H,L_2),\\
 \frac{3R^2}{2q^2k^2}+\{L_4,L_4,L_2\}-\frac{q^2k^2}{2}L_4^2-12k^2P^+(H,
L_2)+10qk^2P^-(H,L_2)=0.
\end{gather*}
This shows that the symmetry algebra is generated by the symmetries $H$, $L_2$, $L_4$
and is closed under commutation.

\begin{example}\label{example2}
We consider our system for the  case   $k=\frac{1}{ 2}$,  i.e., $p=1$, $q=2$:
\[
V=\frac{\alpha }{ 2\sqrt {s^2_1+s^2_2}(s_1+\sqrt {s^2_1+s^2_2})}
=\frac {\alpha }{4 \sin ^2\theta \cos^2(\frac{1}{ 2}\varphi )},
\]
where we now use the fact that
$\psi =\varphi/2 $. This is a special case of system [S7] in \cite{KKMP}.
We form the functions $\Delta _+$ and $\Delta _-$ as before:
\begin{gather*}
\Delta _+(N)f=(\cos\psi \partial _\psi -(N+1)\sin\psi )(\cos\psi \partial _\psi -(N+2)\sin\psi )\\
\phantom{\Delta _+(N)f=}{} \times
\left(-\partial _\theta +\frac{1}{ 2}\left(N+\frac{1}{ 2}\right)\cot\theta\right)f,\\
\Delta _-(N)f=(\cos\psi \partial _\psi +N\sin\psi )(\cos\psi \partial _\psi +(N-1)\sin\psi )
\left(-\partial _\theta -\frac{1}{ 2}\left(N+\frac{1}{ 2}\right)\cot\theta \right)f.
\end{gather*}
The Hamiltonian becomes
\[H=\partial ^2_\theta +\cot\theta \partial _\theta +\frac{1}{4\sin^2\theta }
\partial ^2_\psi  + \frac{\alpha }{ 4\sin^2\theta \cos^2\psi }.\]
We obtain
\begin{gather*}
L_4=-2\{J_1,J_3\}+\alpha \cot\theta \tan^2\psi=-2\{J_1,J_3\}+\alpha \frac{s_3\big(2s^2_1+s^2_2-2s_1\sqrt{s_1^2+s_2^2}\big)}
{s_2^2\sqrt{s_1^2+s_2^2}}\\
L_3=-\frac{4}{ 3}\{J_3,J_3,J_2\}-L_4+2\cot\theta \left(\alpha \tan\psi +\frac{1}{ 3}
\sin\psi \cos\psi \right)\partial _\psi\\
\phantom{L_3=}{} +2\left(\alpha \tan^2\psi +\frac{1}{ 3}(1-2\cos2\psi )\right)
\partial _\theta+\alpha \cot\theta \left(\frac{2}{\cos ^2\psi } -1\right).
\end{gather*}
The operator describing the separation in polar coordinates is
\[L_2=\frac14\left(\partial ^2_\psi +\frac{\alpha }{ \cos ^2\psi }\right),\]
which corresponds to eigenvalue $-\frac14(N+\frac{1}{ 2})^2$. The basic commutation
relation is  $L_3+L_4=[L_4,L_2]$.
\end{example}

\section{Another system}\label{system2}

 We can further investigate
these ideas for the case of the special potential
\begin{gather}\label{ex2}
V=\alpha  \frac{(x+iy)^6}{ (x^2+y^2)^4},
\end{gather}
real in Minkowski space with coordinates $x_1$, $x_2$ where $x=x_1$, $y=ix_2$.
In polar coordinates the Schr\"odinger equation has the form
\begin{gather*}
\left(\partial ^2_r+\frac{1}{ r}\partial _r+\frac{1}{ r^2}\partial
^2_\theta +
\frac{\alpha }{ r^2}e^{6i\theta }-E\right)\Psi (r,\theta )=0.
\end{gather*}
If we write $E=-\beta ^2$ then  typical separable solutions in polar
coordinates are
\[
C_\Omega (\beta r)C_{\Omega /3}\big(\delta e^{3i\theta }\big),
\]
where $C_\nu (z)$ is a solution of Bessel's equation,
$\Psi _\Omega =R(r)\Theta (\theta )$ and $\alpha =-9\delta ^2$. Indeed
\[\big(\partial ^2_\theta -9\delta ^2e^{6i\theta }+\Omega ^2\big)\Theta (\theta )=0.\]

We now construct operators which induce the transformation $\Omega \rightarrow
\Omega \pm 3$ on
the basis func\-tions~$\Psi _\Omega$. To do this transparently we use the
variable $w=\delta e^{3i\theta }$. We make the observation that
\[\Phi _+=\left(-\partial _r+ \frac{\Omega +2}{ r} \right)\left(-\partial _r+ \frac{\Omega +1}{
r}
\right)\left(-\partial _r+ \frac{\Omega }{ r}\right)\left(-\partial _w+\frac{\Omega }{ 3w}\right): \ \ \Psi
_\Omega
\rightarrow  \beta ^3\Psi _{\Omega +3}.\]
Similarly we note that
\[\Phi _-=\left(\partial _r+ \frac{\Omega -2}{ r} \right)\left(\partial _r+ \frac{\Omega -1}{
r}
\right)\left(\partial _r+ \frac{\Omega }{ r}\right)\left(\partial _w +\frac{\Omega }{ 3w}\right): \ \  \Psi
_\Omega
\rightarrow  \beta ^3\Psi _{\Omega -3}.\]

Clearly
$\Phi =\Phi _++\Phi _-$
is an even function of $\Omega$,  hence interpretable as a pure  dif\/ferential
operator. Similarly the operator $\Phi_+-\Phi_-$ is an odd function of $\Omega$,
so ${\tilde \Phi}=\frac{1}{\Omega}(\Phi_+-\Phi_-)$ is a~pure dif\/ferential
operator. We deduce as
previously that
$[H,\Phi ]=[H,{\tilde \Phi}]=0$,
hence we have constructed a quantum superintegrable system.

We can extend these ideas to consider the potential
\begin{gather*}
V=\alpha \frac {(x+iy)^{k-1}}{ (x-iy)^{k+1}},
\end{gather*}
where $k={p/q}$ and $\alpha =-k^2\delta ^2$. The solutions  have the
form
\[\Psi _\Omega =C_\Omega (\beta r)C_{q\Omega/p}
\big(\delta e^{ip\theta/q }\big).\]
In order to map solutions of a f\/ixed $\beta $ eigenspace into solutions we can
use the
transformations $\Omega \rightarrow \Omega \pm p$. These transformations can be
performed by the dif\/ferential operators
\begin{gather*}
\Phi _+=\left(-\partial _r+
\frac{\Omega +p-1 }{r}\right)\cdots\left(-\partial _r+\frac{\Omega }{ r}\right)\left(-\partial _w+
\frac{q\Omega}{ pw}\right)\cdots\\
\phantom{\Phi _+=}{}\times \left(-\partial _w+\frac{\Omega q}{ wp}+q-1 \right): \ \ \Psi _\Omega
\rightarrow
\beta ^p\Psi _{\Omega +p},
\end{gather*}
and
\begin{gather*}
\Phi _-=\left(\partial _r+
\frac{\Omega +1-p}{ r}\right)\cdots \left(\partial _r+\frac{\Omega }{ r}\right)\left(\partial _w+
\frac{q\Omega }{ pw}\right)\cdots \\
\phantom{\Phi _-=}{} \times\left(\partial _w+\frac{q\Omega }{ pw}+1-q\right): \ \ \Psi _\Omega
\rightarrow
\beta ^p\Psi _{\Omega -p},
\end{gather*}
where $w=\delta \exp(ip\theta/q )$. If we make the transformation
$\Omega \rightarrow -\Omega $ then we see that
$\Phi _-(-\Omega )=(-1)^{p+q}\Phi _+(\Omega )$.

Consequently there are two cases to consider:
\begin{enumerate}\itemsep=0pt
\item[(a)]  $p+q$  even: $\Phi^{(+)}=\Phi _++\Phi _-$ is an
even function of $\Omega $,
 hence is a pure dif\/ferential operator, and
$\Phi^{(-)}=\frac{1}{ \Omega }(\Phi _+-\Phi _-)$ is also a pure
dif\/ferential operator.
\item[(b)]  $p+q$  odd: $\Phi^{(+)}=\Phi _+-\Phi _-$ is an even function of
$\Omega $, hence
is a pure dif\/ferential operator, and $\Phi^{(-)}=\frac{1}{ \Omega }(\Phi _++\Phi
_-)$ is also a
pure dif\/ferential operator.
\end{enumerate}
Thus we have superintegrability in both cases.

Further, we can prove the f\/inite closure of the symmetry algebra and  construct
the structure equations.  We write
\[L_3=\Phi ^{(+)}, \qquad L_4=\Phi ^{(-)}\]
 and let  $L_2$ be the dif\/ferential
operator whose eigenvalue
corresponds to $\Omega ^2$, i.e.
\[L_2\Theta (\theta )=\left(-\partial ^2_\theta +\frac{p^2}{ q^2}\delta ^2e^{6i\theta
}
\right)\Theta (\theta )=\Omega ^2\Theta (\theta ).\]
Then a direct computation verif\/ies that the symmetry algebra structure relations
are
\begin{gather*}
[L_2,L_4]=R,\qquad
 [L_2,R]=2p^2\{L_2,L_4\}-p^4L_4,\qquad [L_4,R]=-2p^2L^2_4,
 \end{gather*}
together with the Casimir
\[R^2-\frac{2p^2}{ 3}\{L_2,L_4,L_4\}+\frac{11}{ 3}p^4L^2_4+16p^2H^{2p}=0.\]
Here, $\{A,B\}=AB+BA$, $\{A,B,C\}$ is the analogous  6-term symmetrizer of 3
operators, and  $R=2pL_3+p^2L_4$.
These relations hold  no matter whether $p+q$ is even or odd.

This shows that the symmetries $H$, $L_2$, $L_4$ generate the symmetry algebra, and
that it closes.

\begin{example}\label{example3}
 Take $k=p=q=1$. Then
\[V= -\frac{\delta ^2}{ (x-iy)^2}=- \frac{\delta ^2e^{2i}}{ r^2}^\theta =-\frac{w^2}{ r^2},\]
and
\[\Phi _+f=\left(-\partial _r+ \frac{\Omega }{ r}\right)\left(-\partial _w+ \frac{\Omega }{ w}\right)f,\qquad
\Phi _-f=\left(\partial _r+ \frac{\Omega }{ r}\right)\left(\partial _w+ \frac{\Omega }{ w}\right)f.\]
The Hamiltonian is
\[
H=\partial ^2_x+\partial ^2_y-\frac{\delta ^2}{ (x-iy)^2}.
\]
 This corresponds to system [E14] in \cite{KKMP}.
We f\/ind
\[L_4=-\frac{2}{ \delta }(\partial _x-i\partial _y),\qquad
L_3=-\frac{i}{ \delta }\{x\partial _y-y\partial _x,\partial _x-i\partial _y\}+\frac{1
}{\delta }(\partial _x-i\partial _y)+2\frac {\delta ^2w}{ r}\]
and the constant describing the separation of variables in polar coordinates is
\[L_2=w^2\partial ^2_w+w\partial _w+w^2=
-(x\partial _y-y\partial _x)^2+\delta ^2\frac{x+iy}{ x-iy},\]
which corresponds to the eigenvalue $\Omega ^2$. We also have the structure
relation
$[L_2,L_4]=2L_3+L_4$.
\end{example}
\begin{example}\label{example4}
 Take $k=p=2$, $q=1$ and $w=\delta e^{2i\theta}$.  Then
\[V= -4\delta ^2 \frac{(x+iy)}{ (x-iy)^3}= - 4\frac{\delta ^2e^{4i\theta}}{ r^2},\]
and
\begin{gather*}
\Phi _+f=\left(-\partial _r+ \frac{\Omega +1}{ r}\right)\left(-\partial _r+
\frac{\Omega }{ r}\right)\left(-\partial _w+ \frac{\Omega }{ 2w}\right)f,
\\
 \Phi _-f=\left(\partial _r+ \frac{\Omega -1}{ r}\right)\left(\partial _r+
\frac{\Omega }{ r}\right)\left(\partial _w+\frac {\Omega }{ 2w}\right)f.
\end{gather*}
The Hamiltonian is
\[H=\partial ^2_x+\partial ^2_y-4\delta ^2 \frac{x+iy}{ (x-iy)^3}.\]
 This corresponds to system [E8] in \cite{KKMP}.
We f\/ind
\begin{gather*}
L_3=-\frac{i}{ 6\delta }
\{x\partial _y-y\partial _x,\partial _x-i\partial _y,\partial _x-i\partial _y\}
+L_4-4 \frac{w}{ r^2}(2(r\partial _r+w\partial _w)+1,\\
 L_4= \frac{1}{ \delta }(\partial _x-i\partial _y)^2+4 \frac{w}{ r^2},
 \end{gather*}
and the constant describing the separation of variables in polar coordinates is
\[
L_2=4\big(w^2\partial ^2_w+w\partial _w+w^2\big)=-(x\partial _y-y\partial _x)^2+4\delta ^2\left(\frac{x+iy}{ x-iy}\right)^2,
\]
which corresponds to the eigenvalue $\Omega ^2$. We also have the structure
relation $[L_2,L_4]=4(L_3+L_4)$.
\end{example}

\section{The caged anisotropic oscillator revisited}

In \cite{KKM10b} we introduced
the recurrence relation
method by sketching a proof that the caged anisotropic oscillator was quantum
superintegrable. Here we will provide more details
and show that the symmetry algebra always closes. This result is not new \cite{Marquette20101} but
we include it here to illustrate explicitly how it is obtainable from recurrences obeyed by Laguerre functions.The
system is
\begin{gather*}
H=\partial ^2_x+\partial ^2_y-\mu_1 ^2x^2-\mu_2
^2y^2+
\frac{\frac14-a_1^2}{x^2} + \frac{\frac14-a_2^2}{ y^2},
\end{gather*}
where $\mu _1=p\mu $ and $\mu _2=q\mu $ and $p$, $q$ are positive integers that
we assume are relatively prime.

We look for eigenfunctions for the equation $H \Psi=\lambda
\Psi$  of the form $ \Psi = X Y$ and f\/ind the
normalized solutions
\[ X_n=e^{-\frac12\mu _1x^2}x^{a_1+\frac12} L^{a_1}_n\big(\mu_1x^2\big),\qquad
 Y_m=e^{-\frac12\mu_2y^2}y^{a_2+\frac12} L^{a_2}_m\big(\mu _2y^2\big),\]
where the $L_n^\alpha(x)$ are associated Laguerre functions \cite{AAR}.
Separation in Cartesian coordinates is determined by either of
the symmetry operators
\[L_1=\partial_x^2-\mu_1^2x^2+\frac{\frac14-a_1^2}{x^2},\qquad
L_2=\partial_y^2-\mu_1^2y^2+\frac{\frac14-a_2^2}{y^2},\]
where $H=L_1+L_2$.  For the separated solutions given above we have the
eigenvalue equations
$L_1 \Psi=\lambda_x  \Psi$, $L_2 \Psi=\lambda_y
\Psi$, where
\[\lambda _x=-2\mu _1(2n+a_1+1),\qquad
\lambda _y=-2\mu _2(2m+a_2+1).\]
Thus $H \Psi=E \Psi$ where the energy eigenvalue is
\[E=-2\mu (pn+qm+pa_1+p+qa_2+q).\]
In the foregoing we impose no boundary conditions and the Laguerre functions are
stand-ins for either of the two linearly independent
 solutions of the  second order ordinary eigenvalue equations. In particular,
$n$, $m$ are allowed to be complex.
In order that the eigenspace of~$H$ with eigenvalue $E$ remain invariant under
the action of a recurrence operator that
changes~$m$ and~$n$ we must keep $pn+qm$  constant. One possibility that
suggests itself is
that $n\rightarrow n+q$, $m\rightarrow m-p$. A second possibility is
$n\rightarrow n-q$, $m\rightarrow m+p$.

Now note the recurrence formulas for Laguerre functions (or conf\/luent
hypergeometric functions) viz
\[z\frac{d}{ dz}L^\alpha _p(z)=pL^\alpha _p(z)-(p+\alpha )L^\alpha
_{p-1}(z)=(p+1)L^\alpha _{p+1}(z)-(p+1+\alpha -z)L^\alpha _p(z).\]
We apply  these for the cases $z=\mu_1x^2$, $z=\mu_2 y^2$ and make use of the eigenvalue equations for~$\lambda_x$,~$\lambda_y$, respectively.
Again, we can choose two linearly independent solutions for each eigenvalue
equation, each of which satisf\/ies this recurrence. Then, considering
the symmetry operators as acting on basis functions $\Psi_n=X_nY_n$,
 we have the recurrences
\begin{gather}\label{cagedrecurrence1} D^+(\mu _1,x)X_n=\left(\partial ^2_x-2x\mu _1\partial
_x-\mu _1+\mu ^2_1x^2 +\frac{\frac14
-a_1^2}{ x^2}\right)X_n=-4\mu _1(n+1)X_{n+1},\\
 \label{cagedrecurrence2} D^-(\mu _1,x)X_n=\left(\partial ^2_x+2x\mu _1\partial
_x+\mu _1+\mu ^2_1x^2+ \frac{\frac14
-a_1^2}{ x^2}\right)X_n=-4\mu _1(n+a_1)X_{n-1},\\
 \label{cagedrecurrence3} D^+(\mu _2,y)Y_m=\left(\partial ^2_y-2y\mu _2\partial
_y-\mu _2+\mu ^2_2y^2 +\frac{\frac14
-a_2^2}{ y^2}\right)Y_m=-4\mu _2(m+1)Y_{m+1},\\
 \label{cagedrecurrence4} D^-(\mu _2,y)Y_m=\left(\partial ^2_y+2y\mu _2\partial
_y+\mu _2+\mu ^2_2y^2+ \frac{\frac14
-a_2^2}{ y^2}\right)Y_m=-4\mu _2(m+a_2)Y_{m-1}
\end{gather}
for either basis solution.
Writing $u=m+kn$ where $k=p/q$, we have $m=u-kn$ and we can characterize a
formal eigenfunction corresponding to
energy $E=-2\mu (qu+pa_1+p+qa_2+q)$ by $\Psi_n=X_nY_m$. Let $\Phi^+=D^+(\mu_1
,x)^qD^-(\mu_2 ,y)^p$ and
$\Phi^-=D^-(\mu_1,x)^qD^+(\mu_2,y)^p$. By direct calculation, using recurrences
(\ref{cagedrecurrence1})--(\ref{cagedrecurrence4}), we verify the relations
\begin{gather}\label{cagedraising}
\Phi^+\Psi_n=(-4\mu_1)^q(4\mu_2)^p(n+1)_q(-u+kn-a_2)_p\Psi_{n+q},\\
\label{cagedlowering}
\Phi^-\Psi_n=(4\mu_1)^q(-4\mu_2)^p(-n-a_1)_q(u-kn+1)_p\Psi_{n-q}.
\end{gather}
(Note that this system is very simple to analyze compared to the other systems we study in this paper because these two operators are def\/ined independent of $n$. Hence by the
arguments that we have given for previous examples,
each is a symmetry for the caged oscillator system that is algebraically
independent of the pair $L_1$, $L_2$. No  symmetrization or antisymmetrization is needed.)  Thus the system is
superintegrable. Now we construct the operators
\begin{gather*}
\Phi_1(n) =\Phi^+(n-q)\Phi^-(n):   \\
 \Psi_n\longrightarrow (-16\mu_1^2)^q(-16\mu_2^2)^p(n-q+1)_q
\left(-u+k(n-q)-a_2)_p(-n-a_1)_q(u-kn+1\right)_p\Psi_n,\nonumber\\
 \Phi_2(n) =\Phi^-(n+q)\Phi^+(n): \\
  \Psi_n\longrightarrow (-16\mu_1^2)^q(-16\mu_2^2)^p(-n-q-a_1)_q
\left(u-k(n+q)+1)_p(n+1\right)_q(-u+kn-a_2)_p\Psi_n.\nonumber
\end{gather*}
Though we have indicated a dependence of operators $\Phi_1$, $\Phi_2$ on $n$ in
order to compute their action on a formal
eigenbasis, in fact we see from relations
(\ref{cagedrecurrence1})--(\ref{cagedrecurrence4}) that they are pure
dif\/ferential operators, independent of
the parameter $n$. Further they commute with both $L_1$ and $L_2$. Hence by an
argument that we have given for a
previous example, they must be polynomials in the symmetries $L_1$ and $H$.
Making the replacements
\[ u\leftrightarrow \frac{H+2\mu (pa_1+p+qa_2+q)}{2\mu q},\qquad n\leftrightarrow
\frac{L_1 -2\mu _1(a_1+1)}{4\mu_1},\]
we f\/ind
\begin{gather*}
\Phi_1=\big({-}16\mu_1^2\big)^q(-16\mu_2^2)^p\left(\frac{L_1 -2\mu
_1(a_1+1)}{4\mu_1}  -q+1\right)_q \nonumber \\
\phantom{\Phi_1=}{} \times\left(-\frac{H+2\mu (pa_1+p+qa_2+q)}{2\mu q}+k\left(\frac{L_1 -2\mu
_1(a_1+1)}{4\mu_1}-q\right)-\alpha_2\right)_p\nonumber\\
\phantom{\Phi_1=}{} \times
 \left(-\frac{L_1 -2\mu _1(a_1+1)}{4\mu_1}-\alpha_1\right)_q \nonumber\\
\phantom{\Phi_1=}{} \times \left(\frac{H+2\mu (pa_1+p+qa_2+q)}{2\mu q}-k \frac{L_1 -2\mu
_1(a_1+1)}{4\mu_1}+1\right)_p=P_1(H,L_1)\Psi_n, \\ 
 \Phi_2 =\big({-}16\mu_1^2\big)^q(-16\mu_2^2)^p\left(-\frac{L_1 -2\mu
_1(a_1+1)}{4\mu_1}-q-\alpha_1\right)_q\nonumber\\
\phantom{\Phi_2 =}{} \times
 \left( \frac{H+2\mu (pa_1+p+qa_2+q)}{2\mu q}-k\left(\frac{L_1 -2\mu
_1(a_1+1)}{4\mu_1}+q\right)+1\right)_p\nonumber\\
\phantom{\Phi_2 =}{}
\times \left(\frac{L_1 -2\mu _1(a_1+1)}{4\mu_1}+1\right)_q \\ 
\phantom{\Phi_2 =}{}
\times \left(-\frac{H+2\mu (pa_1+p+qa_2+q)}{2\mu q}+k\frac{L_1 -2\mu
_1(a_1+1)}{4\mu_1}-\alpha_2\right)_p\Psi_n =P_2(H,L_1)\Psi_n.\nonumber
\end{gather*}
The constant terms in the expansions of these relations should be interpreted as
the constants times the identity operator.
Using these results and the eigenvalue formulas $L_1\Psi_n=-2\mu
_1(2n+a_1+1)\Psi_n$, $L_2\Psi_n=-2\mu _2(2u-2kn+a_2+1)\Psi_n$
we can derive the structure equations for the symmetry algebra by acting on a
formal eigenbasis. Then we can use
our previous argument to show that the structure equations
must hold independent of basis. Let $L_3=\Phi^++\Phi^-$ and $L_4=\Phi^+-\Phi^-$.
Then we have
\begin{gather*}
[L_1,L_3]=-4\mu pq L_4,\qquad [L_1,L_4]=-4\mu
pqL_3,\\
[L_3,L_4]=-2P_1(H,L_1)+2P_2(H,L_1),\qquad
L_3^2=L_4^2+2P_1(H,L_1)+2P_2(H,L_1).
\end{gather*}
Thus the symmetry algebra closes. We can take the symmetries $H$, $L_1$, $L_3$ as the
generators with $R=[L_1,L_3] =-4\mu pq L_4$
and rewrite the structure equations as
\begin{gather*}
[L_1,L_3]=R,\qquad [L_1,R]=16\mu^2p^2q^2 L_3,\qquad
 [L_3,R]=8\mu pq P_1(H,L_1)-8\mu pq P_2(H,L_1),\nonumber\\
\frac{1}{16\mu^2 p^2q^2}R^2=L_3^2-2P_1(H,L_1)-2P_2(H,L_1).
\end{gather*}

\begin{example}\label{example5}
 We take the case of
equal frequencies: $\mu _1=\mu _2=\mu$, so $p=q=1$. The correspon\-ding recurrence operators  are
\begin{gather*}
\Phi^+f=\left(\partial ^2_x-2x\mu \partial _x-\mu +\mu ^2x^2+\frac{A_1}{ x^2}\right)\left(
\partial ^2_y+2y\mu \partial _y+\mu +\mu ^2y^2+\frac{A_2}{ y^2}\right)f,\\
\Phi^-f=\left(\partial ^2_y-2y\mu \partial _y-\mu +\mu ^2y^2+\frac{A_2}{ y^2}\right)\left(
\partial ^2_x+2x\mu \partial _x+\mu +\mu ^2x^2+\frac{A_1}{ x^2}\right)f,
\end{gather*}
where $A_j=\frac14-a_j^2$.
To proceed we need to have available the following operators:
\begin{gather*}
L_1=\partial ^2_x-\mu ^2x^2+ \frac{A_1}{ x^2},\qquad L_2=\partial ^2_y-\mu ^2y^2+ \frac{A_2}{ y^2},\qquad
M=(x\partial _y-y\partial _x)^2+ A_1 \frac{y^2}{ x^2} + A_2 \frac{x^2}{ y^2},
\end{gather*}
where $L_1$, $L_2$ come from our general theory and we recall that $H=L_1+L_2$. We  calculate the symmetry
operators that our method implies.
\[L_3=\Phi _++\Phi _-=2L_1L_2+4\mu ^2M-2\mu ^2,\qquad
L_4=\Phi _+-\Phi _-=\mu [M,L_1].\]
Then we have
\[[L_1,L_3]=-4\mu L_4,\qquad [L_1,L_4]=-4\mu L_3.\]
This method implies the existence of the symmetry operator $M$ from the expression for $L_3$.
\end{example}

\section{The TTW system}

A similar but more complicated procedure works for the  quantum TTW system
\cite{TTW, TTW2}. Here  the
Hamiltonian is
\begin{gather*}
H=\partial ^2_r+\frac{1}{ r}\partial _r-\omega^2 r^2+\frac{1}{
r^2}\left(\partial ^2_\theta +
\frac{\alpha }{ \sin ^2(k\theta )} + \frac {\beta }{ \cos ^2(k\theta )}\right),
\end{gather*}
where we take $k=\frac{p}{ q}$ as before. The general solution of the eigenvalue
problem $H\Psi =E\Psi $ is
\begin{gather}\label{sepeigenfunctions}
\Psi  = e^{-\frac{\omega }{ 2}r^2}r^{k(2n+a+b+1)}
L^{k(2n+a+b+1)}_m(\omega r^2)
(\sin (k\theta ))^{a+\frac{1}{ 2}}(\cos (k\theta ))^{b+\frac{1}{ 2}}
P^{a,b}_n(\cos (2k\theta )),\end{gather}
where we have taken $\alpha =k^2(\frac{1}{ 4} -a^2)$ and $\beta =k^2(\frac{1}{
4}
-b^2)$. The $L$-functions are associated Laguerre and the $P$-functions are
Jacobi,
not polynomials in general, \cite{AAR}.   We consider the functions
\[\Pi  = e^{-\frac{\omega }{2}r^2}r^{k(2n+a+b+1)}
L^{k(2n+a+b+1)}_m(\omega r^2)P^{a,b}_n(x)=Y^{ A}_m(r)X^{a,b}_n(x),\]
where $x=\cos (2k\theta )$,  $X=P$ or $Q$ and $Y=S$ or $T$. (Here $\Pi$ is obtained from $\Psi$ by a
gauge transformation to remove the angular factors
$(\sin (k\theta ))^{a+\frac{1}{ 2}}(\cos (k\theta ))^{b+\frac{1}{ 2}}$.)
 By this we mean that $P^{a,b}_n(x)$ is a
Jacobi polynomial if $n$ is an integer. Otherwise it is given by its
hypergeometric expression. If $X=Q$ then this denotes the associated second
solution of the Jacobi dif\/ferential equation. Similar remarks apply to the
choice of $Y=S$.  We  have def\/ined this function to be
\[S^{ A}_m(r)=e^{-\frac{\omega }{ 2}r^2}r^{k(2n+a+b+1)}
L^{k(2n+a+b+1)}_m\big(\omega r^2\big)\]
and $T^{ A}_m(r)$ to be a second independent solution. The energy eigenvalue
is given
by
\begin{gather}\label{energyev}
E=-2\omega \left(2(m+nk)+1+(a+b+1)k\right)\end{gather}
and $ A=k(2n+a+b+1)$.
The separation equation for $\Theta (\theta )$ is
\[{\tilde L}_2\Theta=\left(\partial ^2_\theta + \frac{\alpha }{ \sin ^2(k\theta )} +
\frac{\beta }{\cos ^2(k\theta )}\right)\Theta (\theta )=-k^2(2n+a+b+1)^2\Theta (
\theta )=-A^2 \Theta (\theta )\]
and ${\tilde L}_2$ is a symmetry operator for  the system. Under the gauge
transformation~${\tilde L}_2$  goes to a~symmetry
that we shall  call~$L_2$ and which has the same eigenvalues.
We see from the expression for~$E$ that in order that an energy eigenvalue  be
unchanged for dif\/ferent values of~$m$,~$n$
 we must f\/ix $u=m+nk$. The   two transformations
\[
n\rightarrow n+q,\qquad m\rightarrow m-p\qquad {\rm
and}\qquad
n\rightarrow n-q,\qquad m\rightarrow m+p\]
will each achieve this.

Consider the functions $X^{a,b}_n(x)$. If we want to raise or lower
the index $n$ we can do so with the operators~\cite{AAR}
\begin{gather*}
J^+_n X^{a,b}_n(x)= (2n+a+b+2)(1-x^2)\partial
_xX^{a,b}_n(x)\\
\phantom{J^+_n X^{a,b}_n(x)=}{} +(n+a+b+1)(-(2n+a+b+2)x-(a-b))X^{a,b}_n(x) \\
\phantom{J^+_n X^{a,b}_n(x)}{}
 =2(n+1)(n+a+b+1)X^{a,b}_{n+1}(x)
 \end{gather*}
and
\begin{gather*}
J^-_n X^{a,b}_n(x)= -(2n+a+b)(1-x^2)\partial
_xX^{a,b}_n(x)\\
\phantom{J^-_n X^{a,b}_n(x)=}{} -n((2n+a+b)x-(a-b))X^{a,b}_n(x)=2(n+a)(n+b)X^{a,b}_{n-1}(x).
\end{gather*}
Similarly, for the  functions ${\cal Y}^A_m(R)=\omega^{A/2}Y^{A}_m(r)$ where
$R= r^2$ we can deduce the
relations~\cite{AAR}
\begin{gather*}
K^+_{A,m}{\cal Y}^A_m(R)=\left\{(A+1)\partial _R-\frac{E}{ 4 }-\frac{1}{
2R}A(A+1)\right\}{\cal Y}^A_m(R)
= -\omega{\cal Y}^{A+2}_{m-1}(R),\\
 K^-_{A,m}{\cal Y}^A_m(R)=\left\{(-A+1)
\partial _R-\frac{E}{ 4}+\frac{1}{ 2R}A(1-A)\right\}{\cal Y}^A_m(R)
=-\omega(m\!+\!1)(m\!+\!A) {\cal Y}^{A-2}_{m+1}(R) .
\end{gather*}
(Note that $E=-2\omega[2(m+nk)+1+(a+b+1)k]$, $ A=k(2n+a+b+1)$ and, ef\/fectively, the operator
 $K^+$ is lowering $m$ by $1$ and raising $n$ by $q/p$, whereas the operator $K^-$ is
raising~$m$ by~$1$ and lowering $n$ by $q/p$. Here, $E$ is f\/ixed.)
 We now  construct the two operators
\begin{gather}\label{recurrence3} \Xi _+=K^+_{A+2(p-1),m-(p-1)}\cdots
K^+_{A,m}J^+_{n+q-1}\cdots J^+_n\end{gather}
and
\begin{gather}\label{recurrence4}\Xi _-=K^-_{A-2(p-1),m+p-1}\cdots
K^-_{A,m}J^-_{n-q+1}\cdots J^-_n. \end{gather}
When applied to a basis function $\Psi_n={\cal Y}^A_m(R)X^{a,b}_n(x)$ for f\/ixed
$u=m+kn$, (so $m=u-kn$)
these operators  raise and lower indices according to
\begin{gather}\label{raise1} \Xi _+\Psi_n=2^q(-1)^p \omega^p(n+1)_q(n+a+b+1)_q\Psi_{n+q},\\
 \label{lower1} \Xi _-\Psi_n=2^q \omega^p
(-n-a)_q(-n-b)_q(u-kn+1)_p\left(-u-k(n+a+b+1)\right)_p\Psi_{n-q},
\end{gather}
where $(\alpha)_q=(\alpha)(\alpha+1)\cdots(\alpha+q-1)$ for nonnegative integer
$q$, and we note the relation
$(-\alpha)_q=(-1)^q(\alpha-q+1)_q$.
From the explicit expressions (\ref{recurrence3}), (\ref{recurrence4}) for these
operators it is easy to verify that
under the transformation $n \rightarrow  -n-a-b-1$  we have ${\Xi_+} \rightarrow
\Xi_-$ and ${\Xi_-} \rightarrow \Xi_+$.
Thus $\Xi=\Xi_++\Xi_-$  as a polynomial in $n$ and $u$ is unchanged under this
transformation.
Therefore it is a polynomial in $(2n+a+b+1)^2$ and $u$.  As a consequence of
the relation $\lambda =-k^2(2n+a+b+1)^2$,   in the
expansion of~$\Xi$ in terms of powers of $(2n+a+b+1)^2$ and~$E$ we can replace $(2n+a+b+1)^2$
by $L_2/k^2$ and~$E$ by~$H$ everywhere they occur, and express~$\Xi$ as a
pure dif\/ferential operator, independent of parameters.  (Note that in the
expansion of~$\Xi$ in terms of the parameters, a term~$ W E$
 is replaced by~$WH$
with the $W$ operator on the left.)
Clearly this operator, which we will also call~$\Xi$ commutes with~$H$ on the
eigenspaces of $H$. However, the same argument as used in
Example~\ref{example1} shows that in fact~$\Xi$ commutes with~$H$ in general,
thus it is a symmetry operator for~$H$.

We can also easily see that under the transformation $n \rightarrow  -n-a-b-1$
the operator $\Xi_+ -\Xi_-$ changes sign, hence the operator
 ${\tilde \Xi}=(1/(2n+a+b+1))(\Xi_+ -\Xi_-)$ is unchanged under this
transformation. Again, making the replacements $(2n+a+b+1)^2$
by $L_2/k^2$ and $u$ by
$-2 -(H+2\omega k(a+b+1))/4\omega$ we can express~${\tilde \Xi}$ as a pure
dif\/ferential operator, independent of parameters,
and it is a symmetry operator for~$H$. Each of these symmetries has a nonzero
commutator with~$L_2$, so each is
algebraically independent of the set $H$, $L_2$. This proves that the TTW system is
quantum superintegrable for all rational~$k$.
We set $L_3= \Xi$, $L_4={\tilde \Xi}$.

Using the explicit relations (\ref{raise1}), (\ref{lower1}) for the action of
the raising and lowering opera\-tors~$\Xi_\pm$ on a basis
we can obtain very detailed information about the structure of the symmetry
algebra generated by $L_2$, $L_3$, $L_4$.
Applying the raising operator to a basis function, followed by the lowering
operator, we obtain the result
\begin{gather} \Xi_-(n+q)\Xi_+(n)\Psi_n
=(-1)^p4^q\omega^{2p}(n+1)_q(n+a+1)_q(n+b+1)_q(n+a+b+1)_q\nonumber\\
\phantom{\Xi_-(n+q)\Xi_+(n)\Psi_n =}{}
\times(-u+kn)_p\left(u+k(n+a+b+1)
+1\right)_p\Psi_n =\xi_n\Psi_n.
\label{LR1}
\end{gather}
Reversing the order we f\/ind
\begin{gather} \Xi_+(n-q)\Xi_-(n)\Psi_n
 =(-1)^p4^q\omega^{2p}(-n)_q(-n-a)_q(-n-b)_q(-n-a-b)_q(u-kn+1)_p\nonumber\\
 \phantom{\Xi_+(n-q)\Xi_-(n)\Psi_n =}{}\times\left(-u-k(n+a+b+1)\right)_p\Psi_n
  =\eta_n\Psi_n.\label{RL1}
\end{gather}
Thus the action of the operator $\Xi_-(n+q)\Xi_+(n)+ \Xi_+(n-q)\Xi_-(n)$ on any
basis function $\Psi_n$ is to multiply it by
$\xi_n+\eta_n$. However, it is easy to check from expressions (\ref{raise1}),
(\ref{lower1}) and from~(\ref{RL1}),~(\ref{LR1})
that under the transformation $n \rightarrow  -n-a-b-1$  we have
$\Xi_-(n+q)\Xi_+(n)\leftrightarrow \Xi_+(n-q)\Xi_-(n)$ and
  $\xi_n\leftrightarrow \eta_n$. Thus $\Xi^{(+)}=\Xi_-(n+q)\Xi_+(n)+
\Xi_+(n-q)\Xi_-(n)$ is an even polynomial operator in $(2n+a+b+1)$,
polynomial in $u$, and
 $\xi_n+\eta_n$ is an even polynomial function in $(2n+a+b+1)$, polynomial in
$u$. Furthermore, each of $\Xi_-(n+q)\Xi_+(n)$ and $\Xi_+(n-q)\Xi_-(n)$ is unchanged under the transformation
$u\longrightarrow -u-(a+b+1)-1$, hence each is a polynomial of order $p$ in
$\left(2u+(a+b+1)k+1\right)^2=E^2/4\omega^2$. Due to the multiplicative factor~$\omega^{2p}$ in each of these expressions
we conclude that $\Xi^{(+)}$ is a
symmetry operator whose action on a basis is
 given by a~polynomial operator $P^{(+)}(H^2,L_2,\omega^2,a,b)$. In fact,
\[ \Xi^{(+)}=P^{(+)}\big(H^2,L_2,\omega^2,a,b\big).\]
The proof is analogous to that given in Example~\ref{example1}. We write the
operator $\Xi^{(+)}-P^{(+)}$ in canonical form
$A\partial^2_{xR}+b\partial_x+C\partial_R+D$. Then on an arbitrary basis we have
\[\left(A\partial^2 _{xR}+B\partial _x+C\partial _R+D\right)\Psi_n=0.\]
It follows via the usual Wronskian argument that $A=B=C=D=0$.

Similarly we note that the operator $\Xi^{(-)}=(\Xi_-(n+q)\Xi_+(n)-
\Xi_+(n-q)\Xi_-(n))/(2n+a+b+1)$ is an even polynomial
in $(2n+a+b+1)$, as is $(\xi_n-\eta_n)/(2n+a+b+1)$. Also it is polynomial in $E^2$ and $\omega^2$. Thus
$\Xi^{(-)}=P^{(-)}(H^2,L_2,\omega^2,a,b)$ is a symmetry operator
which is a polynomial function of all of its variables.

Now we can compute the structure relations explicitly by evaluating the
operators on an eigenfunction basis. As we have demonstrated,
these relations must then hold everywhere. The results are:
\begin{gather*}
 [L_2,L_4]=-4k^2qL_3-4k^2q^2L_4,\\
  [L_2,L_3]=2q\{L_2,L_4\}+4k^2q^2L_3+8k^2q^3L_4,\\
 [L_3,L_4]=2qL_4^2-2P^{(-)}\big(H^2,L_2,\omega^2,a,b\big),\\
6k^2L_3^2+\{L_2,L_4,L_4\}+6k^2q\{L_3,L_4\}+28k^2q^2L_4^2-4k^2qP^{(-)}\big(H^2,
L_2,\omega^2,a,b\big)\nonumber\\
 \qquad{} -12k^2P^{(+)}\big(H^2,L_2,\omega^2,a,b\big)=0. 
\end{gather*}
Here, $\{A,B\}=AB+BA$ and $\{A,B,C\}$ is the analogous  6-term symmetrizer of 3
operators.
A~more transparent realization, for $R=-4k^2qL_3-4k^2q^2L_4$, is
\begin{gather*}
[L_2,L_4]=R,\\
 \label{[L2,R]} [L_2,R]=-8k^2q^2\{L_2,L_4\}-16k^4q^4L_4,\\
 [L_4,R]=8k^2q^2L_4^2-8k^2qP^{(-)}\big(H^2,L_2,\omega^2,a,b\big),\\
\frac{3}{8k^2q^2}R^2+22k^2q^2L_4^2+\{L_2,L_4,L_4\}-4k^2qP^{(-)}\big(H^2,L_2,\omega^2,a,b\big)\nonumber\\
\qquad{} -12k^2P^{
(+)}\big(H^2,L_2,\omega^2,a,b\big)=0. 
\end{gather*}
From this realization we see that the symmetries $H$, $L_2$, $L_4$ generate a closed
symmetry algebra.

\begin{example}\label{example6}
 We consider the TTW system with $k=p=q=1$. This is essentially the same as Example~\ref{example5}
but with dif\/ferent assumptions. We choose the operators $X=\partial_x^2-\omega^2+A_1/x^2$, $Y=\partial_y^2-\omega^2+A_2/y^2$ and $L_2=M$ from  the
previous example where now $A_1=\frac14-a^2$, $A_2=\frac14-b^2$. Then
\[\Xi _+f=J^+_nK^+_{A,m}f,\qquad \Xi _-f=J^-_nK^-_{A,m}f\]
where $A=2n+a+b+1$, and we form the usual combinations
\[L_3=\Xi _++\Xi _-=-\frac{1}{ 8\omega } \{L_2,X-Y\}+\frac{1}{ 4\omega }
\big(a^2-b^2\big)H-\frac{1}{ 4\omega }(X-Y)+\frac{1}{ 8\omega } [M,X]\]
and
\[L_4=\frac{1}{ A} (\Xi _+-\Xi _-)=\frac{1}{ 8\omega } [X,M].\]
From these expressions we deduce that
$[L_2,L_4]=-4(L_3+L_4)$, as expected.     Note that it follows from these equations that the second order operator $X-Y$ is a symmetry.
Indeed,  taking the commutator of $H$ with $L_3$ we f\/ind $[H,X-Y]L_2+L_2[H,X-Y]=0$.  From this and formal adjoint properties one can conclude that $[H,X-Y]=0$, so $X-Y$ is a 2nd order symmetry. However, $Y-Y$ doesn't belong to the algebra we have already found. In other words our standard procedure didn't f\/ind the lowest order generator for the symmetry algebra in this case; it determined a proper subalgebra of the full symmetry structure.  To f\/ind the full algebra we need to check that the symmetry algebra generated by $X-Y$, $L_2$, $H$  closes at f\/inite order, though in this particular example we already know this to be the case. In the next subsection we will give another approach to this problem that describes how the missing symmetry can always be found and shows that it expressible in terms of the fundamental raising and lowering operators.
\end{example}

\subsection[The symmetry $L_5$]{The symmetry $\boldsymbol{L_5}$}

We investigate the fact that, as shown in Example~\ref{example6}, our method doesn't always give generators of minimal order.
 For the case $k=1$ we know from the structure theory of 2nd order superintegrable 2D systems with nondegenerate potential~\cite{KKM20041,KKM20061} that the space of 3rd order symmetries is 1-dimensional. Thus, we know that there is a 2nd order symmetry
 operator $L_5$ for this case, independent of $L_2$, $H$, such that $[L_2,L_5]=L_4$. We will show how to obtain this symmetry from the raising and lowering operators
 $\Xi_\pm$, without making use of multiseparability. Thus, for general rational $k$ we look for a symmetry operator $L_5$ such that $[L_2,L_5]=L_4$.
 Applying this condition to a formal eigenbasis of functions $\Psi_n$ we obtain the result
\[L_2(L_5\Psi_n)=-k^2(2n+a+b+1)^2(L_5\Psi_n)+\frac{\Xi_+-\Xi_-}{2n+a+b+1}\Psi_n.\]
The general solution is
\begin{gather}
L_5=-\frac{1}{4qk^2}\bigg(\frac{\Xi_+}{(2n+q+a+b+1)(2n+a+b+1)}\nonumber\\
\phantom{L_5=}{}+\frac{\Xi_-}{(2n-q+a+b+1)(2n+a+b+1)}\bigg)+\beta_n,\label{L51}
\end{gather}
where $\beta_n$ is a rational scalar function. It is easy to check that the quantity in parentheses is a rational scalar function of $(2n+a+b+1)^2$.
Thus we will have a true constant of the motion, polynomial in the momenta, provided we can choose $\beta_n$ such that the full quantity~(\ref{L51})
 is polynomial in $(2n+a+b+1)^2$. To determine the possibilities we need to investigate the singularities of this solution  at
$n=-\frac{a+b+1}{2}$, $n=-\frac{\pm q+a+b+1}{2}$. It is easy to check that the operator has a removable singularity at $n=-\frac{a+b+1}{2}$.
 In the special case $k=1$ the singularities at $ n=-\frac{\pm 1+a+b+1}{2}$
are removable, provided we set $\beta_n=-\frac{H(a^2-b^2)}{32(2n+a+b+2)(2n+a+b)}$. Further, the expression for $L_5$ in this case $k=1$ implies
\[\{L_5,L_2\}=-2L_5+\frac12(L_3+L_4)+\frac{H}{16}\big(a^2-b^2\big).\]
 Similar calculations show that the symmetries $L_2$, $L_5$, $H$ generate the full
closed symmetry algebra, which contains our original symmetry algebra properly.  For general rational $k$ we can set
\[\beta_n=-\frac{Q(H)}{4qk^2(2n+q+a+b+1)(2n-q+a+b+1)}\]
and determine a polynomial  $Q(H)$ such that the operator has a removable singularity at $n=-(q+a+b+1)/2$, i.e., such that the residue is~$0$. (Since the solution is a polynomial in $(2n+a+b+1)^2$ we don't have to worry about the singularity at $n=-(-q+a+b+1)/2$.) Fixing $E$ we can consider the operators $K^\pm$ as depending on~$A$ alone. Note that
\begin{gather*}
 K^+_{-1}=-\frac{E}{4}=-\frac{H}{4},\qquad J^+_{-1-(a+b)/2}=\frac12\big(a^2-b^2\big),\qquad K^+_A=K^-_{-A},\qquad J^-_n=J^+_{-n-a-b-1}.
 \end{gather*}
Setting $\mu=-(a+b+1)/2$, we see that we have to evaluate the product $\Xi_+\Psi_{-q/2-\mu}$, which factors as
\[
\Xi_+\Psi_{-q/2+\mu}=\left(K^+_{p-2}\cdots K^+_{-p+2}K^+_{-p}{\cal Y}_{-p}\right)\left( J^+_{q/2+\mu-1}\cdots J^+_{-q/2+\mu+1}J^+_{-q/2+\mu}X_{-q/2+\mu}\right).
\]
Thus to compute the residue  we can pair up terms using the following consequences of the recurrence relations above:
\begin{gather*}
J^-_{n+1}J^+_n=J^+_{-n-a-b-2}J^+_n   \sim 4(n+1)(n+a+b+1)(n+a+1)(n+b+1),\\
 K^-_{A+2}K^+_A=K^+_{-A-2}K^+_A \sim -\frac{\omega^2}{4}\left(\frac{H}{2\omega}+1+A\right)\left(\frac{H}{2\omega}-1-A\right).
 \end{gather*}
For example, consider the $y$-dependent terms with $p$ odd. Then the central term in the $K$-factors is multiplication by the constant  $K^+_{-1}=-\frac{H}{4}$. Note that the operators on either side of $K^+_{-1}$ pair up to give $ K^+_{1}K^+_{-3}$ which acts as multiplication by a constant. Then we consider the next pair $K^+_3K^+_{-5}$, and so on to evaluate the product. If~$p$ is even there is no central term and we start from the inside by pairing $K^+_0K^+_{-2}$, followed by $K^+_2K^+_{-4}$, and so on. The same procedure works for the $J$-factors, except that the step size is~$1$, rather than~$2$.

The full computation is simple in principle, but technical. It breaks up into 3 cases:

\medskip

\noindent $ p$  even, $q$  odd:
\begin{gather*}
Q(H)= \big(a^2-b^2\big)\prod_{\ell=1}^{p/2} \left(\big({-}\omega^2\big)\left(\frac{H}{4\omega}+\frac12-\ell\right)
\left(\frac{H}{4\omega}-\frac12+\ell\right)\right)  \\
\phantom{Q(H)=}{} \times\prod_{s=1}^{(q-1)/2} \left(\frac14 (-a-b+2s)(a+b+2s)(a-b+2s)(-a+b+2s)\right),
\end{gather*}
$ p$  odd, $q$  odd:
\begin{gather*}
Q(H)= -\frac{H(a^2-b^2)}{4}\prod_{\ell=1}^{(p-1)/2}\left(\big({-}\omega^2\big)\left(\frac{H}{4\omega}-\ell\right)\left(\frac{H}{4\omega}+\ell\right)\right)\\
\phantom{Q(H)=}{} \times
 \prod_{s=1}^{(q-1)/2}\left(\frac14(-a-b+2s)(a+b+2s)(a-b+2s)(-a+b+2s)\right),
 \end{gather*}
$ p$  odd, $q$  even:
\begin{gather*}
Q(H)= -\frac{H}{2}\prod_{\ell=1}^{(p-1)/2}\left(\big({-}\omega^2\big)\left(\frac{H}{4\omega}-\ell\right)\left(\frac{H}{4\omega}+\ell\right)\right)\\
\phantom{Q(H)=}{} \times
 \prod_{h=0}^{q/2-1}\bigg(\frac14(-q-a-b+2h+1)(-q+a+b+2h+1)(-q+a-b+2h+1)\\
\phantom{Q(H)=}{} \times
 (-q-a+b+2h+1)\bigg).
 \end{gather*}

Again, by construction one can check that, for general rational $k$, the operators  $L_2$, $L_5$, $H$ generate a symmetry algebra that properly contains our original algebra and that closes. The new structure relations are somewhat more complicated than before, but computable.
Our basic point is that the full set of symmetries is
 generated by the fundamental raising and lowering operators~$\Xi_\pm$, even though they themselves are not polynomial symmetries.

\section{One-variable models}

The recurrence operators that we have introduced via special function theory
lead almost immediately
to one-variable function space models
that represent the symmetry algebra action in terms of dif\/ference operators on
spaces of polynomials. We will
 illustrate this for two systems, the caged anisotropic oscillator and the TTW
system, and will describe some of
 the information that can be gleaned from such models.

  {\bf The caged anisotropic oscillator.}  From expressions
(\ref{cagedraising}), (\ref{cagedlowering}) and the eigenvalue equations for $H$
and $L_1$ it is easy to write down a function space model for irreducible
representations of the symmetry algebra. Note that since~$H$ commutes with all
elements of the algebra, it corresponds to multiplication by a constant~$E$ in
the model. We let the complex variable~$t$ correspond to~$n$. Then the action of
the symmetry algebra on the  space of polynomials $f(t)$ is given by dif\/ference
operators
\begin{gather}\label{cagedraisingt}
\Phi^+f(t)=(-4\mu_1)^q(4\mu_2)^p(t+1)_q(-u+kt-a_2)_pf(t+q),\\
 \label{cagedloweringt}
\Phi^-f(t)=(-4\mu_1)^q(4\mu_2)^p(-t-a_1)_q(-u-kt+1)_pf(t-q),\\
 \label{cagedeigenvaluest} Ef(t)=-2\mu(qu+pa_1+p+qa_2+q)f(t),\qquad
L_1f(t)=-2\mu_1(2t+a_1+1)f(t),\\
 \mu_1=p\mu,\qquad \mu_2=q\mu.\nonumber
 \end{gather}
These operators satisfy relations
\begin{gather*}
\Phi^+\Phi^-=P_1(E,L_1),\qquad \Phi^-\Phi^+=P_2(E,L_1),\\
[L_2,\Phi^+]=-4pq\mu
\Phi^+,\qquad [L_2,\Phi^-]=4pq\mu \Phi^-.
\end{gather*}
(Note that, by using a similarity transformation via the Mellin transform and its inverse, we could
also transform the model (\ref{cagedraisingt})--(\ref{cagedeigenvaluest})
 into a realization by dif\/ferential operators
in one variable.)
For a f\/inite dimensional irreducible representation the unnormalized
eigenfunctions of $L_2$ are delta
functions $\Psi_{t_N}(t)=\delta(t+t_N)$. To derive an inner product on the
representation space we can require
that the adjoint of $\Phi^+$ is $\Phi^-$ and that $L_2$ is self-adjoint.

We can use the model to f\/ind several families of
f\/inite dimensional and inf\/inite dimensional irreducible representations of the symmetry algebra, only some of which correspond to those
that arise from the  quantum mechanical eigenvalue  associated with real physical systems.
As an example of the use of the model to construct f\/inite dimensional representations let us assume
that $a_1$, $a_2$ are real and  look for a realization such that $t_N=t^0+Nq$ where $N$ is an integer and $0\le t^0<q$ is f\/ixed.
 Let $N_0$ be the smallest integer such that $\Psi_{N_0}=\delta(t+t_{N_0})$ is an eigenfunction, and let $N_1$ be the largest
such integer. Then we must have $\Phi^-\Psi_{N_0}=0$, $\Phi^+\Psi_{N_1}=0$. We see from the model that one way to accomplish
this is to choose
\[
u-k[t^0+N_0q]+1=-p_0,\qquad t^0+N_1q+1=-q_0,
\]
where $p_0$, $q_0$ are integers such that $0\le p_0<p$ and $0\le q_0<q$. Now set $M=N_1-N_0$, so that
the dimension of the representation is $M+1$.
Simple algebra gives the eigenvalue
\[ E=2Mpq\mu -2\mu[p(a_1-q_0)+q(a_2-p_0)]\]
for $H$ and eigenvalues
\[
4\mu pq(N_1-N)-2\mu p(a_1-2q_0-1),\qquad N=N_0,N_0+1,\dots, N_1,
\]
for $L_1$.

 {\bf The TTW potential.}  We use expressions (\ref{raising1}),
(\ref{lowering1}) and the
 eigenvalue equations for~$H$ and~$L_2$  to write def\/ine a function space model
for irreducible  of the symmetry algebra.
The results can be presented in a simpler form if we use a gauge transformation
$\Phi_n=g(n)\Psi _n$ where
\[\frac{g(n)}{g(n-q)}=\frac{(n-q+1)_q}{\left(-u-k(n+a+b+1)\right)_p(-n-a)_q}\]
and  introduce a complex variable $s$, corresponding to $n+(a+b+1)/2$. Then the
action of the symmetry algebra on the
space of polynomials $f(s)$ is given by dif\/ference operators
\begin{gather*}
 L_2\Phi(s)=-4k^2s^2\Phi(s),\qquad  E=-2\omega\left(2u+1+k(a+b+1)\right),\\
L_4\Phi(s)=\frac{2^{q-1}}{s}\Bigg(\left(s+\frac{a-b+1}{2}\right)_q\left(s+\frac{a+b+1}{2}\right)_q\nonumber\\
\phantom{L_4\Phi(s)=}{}\times
\left(u+k\left(s+\frac{a+b+1}{2}\right)+1\right)_p\Phi(s+q)
 -\left (-s+\frac{a-b+1}{2}\right)_q\nonumber\\
\phantom{L_4\Phi(s)=}{}\times
 \left(-s+\frac{a+b+1}{2}\right)_q\left(u+k\left(-s+\frac{a+b+1}{2}\right)
+1\right)_p\Phi(s-q)\Bigg).
\end{gather*}
From these expressions it is straightforward to show that the space of
polynomials in the va\-riab\-le~$s^2$
is invariant under the action of the one-variable dif\/ference operators.
For a f\/inite dimensional irreducible representation the unnormalized
eigenfunctions of $L_2$ are delta
functions $\Phi_{s_N}(t)=\delta(s+s_N)$. To derive an inner product on the
representation space we can require
that~$L_2$ and~$L_4$  are self-adjoint.

We can use the dif\/ference model to f\/ind several families of
f\/inite dimensional and inf\/inite dimensional irreducible representations of the symmetry algebra,
only some of which correspond to those
that arise from the  quantum mechanical eigenvalue  associated with real physical systems.
 As an example of the use of the model to construct f\/inite dimensional representations let us
assume that $a$, $b$, $\omega$ are real and  look for a realization such that $s_N=s^0+Nq$ where $N$ is an integer
and $0\le s^0<q$ is f\/ixed. Let $N_0$ be the smallest integer such that $\Phi_{N_0}=\delta(s+s_{N_0})$ is an eigenfunction,
and let $N_1$ be the largest such integer.  We see from the model that one way to accomplish this is to choose
\[ u+k\left(-s^0-N_0q+\frac{a+b+1}{2}\right)+1=-p_0,\qquad s^0+N_1q+\frac{a+b+1}{2}=-q_0,\]
where $p_0$, $q_0$ are integers such that $0\le p_0<p$ and $0\le q_0<q$. Now set $M=N_1-N_0$, so that the dimension
of the representation is $M+1$.
We f\/ind
\[ E=2\omega\left(2pM+a+b+2kq_0+2p_0+2\right)\]
for $H$ and eigenvalues
\[-4k^2\left((N_1-N)q+\frac{a+b+1}{2}+q_0\right)^2,\qquad N=N_0,N_0+1,\dots, N_1, \]
for $L_2$.

  {\bf The system (\ref{ex2}).} For completeness we give the simple one variable model for this case:
\begin{gather*} 
L_2F(\Omega)=\Omega^2F(\Omega),\qquad  \Phi_+F(\Omega)=\beta^p \left(F(\Omega+p)+(-1)^{p+q}F(\Omega-p)\right),
\\
 \Phi_-F(\Omega)=\frac{1}{\Omega}\beta^p\left(F(\Omega+p)-(-1)^{p+q}F(\Omega-p)\right).\nonumber
 \end{gather*}

\section{St\"ackel transforms and the recurrence method}

The theory of St\"ackel transforms does not guarantee that if two classical
2D superintegrable systems are related by a St\"ackel transform and if one system is quantum superintegrable,
then the other system is also quantum
superintegrable and the quantum systems are related by a~St\'ackel transform. Some additional conditions must be fulf\/illed,
\cite{KMP10}.
However, if one of the systems is known to be quantum superintegrable via our recurrence relation method, then it is automatic that
the second quantum system is also superintegrable and a quantum St\"ackel transform of the f\/irst.  We will give a single example
which makes clear the general proof.

In \cite{PW2010} there was introduced a new family of Hamiltonians with a deformed Kepler--Coulomb potential dependent
 on an indexing parameter $k$ which was shown to be related to the TTW oscillator system system via coupling constant metamorphosis.
The authors showed that this deformed Kepler system is classically superintegrable for all rational $k=p/q$, and in \cite{KKM10b}
we used the canonical equations for higher order symmetry operators to show that
that it is quantum superintegrable. Here we use the recurrence relations obeyed by the eigenfunctions of the separating
symmetry operators to give a new proof of quantum superintegrability and, also, to obtain the structure equations.

As stated above, expressed in polar coordinates $r$, $\theta$, the quantum TTW system is $H\Psi=E\Psi$ or
\begin{gather}\label{TTWham1} \left(\partial ^2_r+\frac{1}{ r}\partial _r+\frac{1}{
r^2}\partial ^2_\theta  -\omega^2 r^2+
\frac{\alpha }{r^2 \sin ^2(k\theta )} + \frac {\beta }{ r^2\cos ^2(k\theta )}\right)\Psi=E\Psi.\end{gather}
The deformed Kepler--Coulomb system, expressed in polar coordinates $R$, $\phi$, is $H'\Psi={\cal E}\Psi$ or
\begin{gather}\label{deformedKepler}
\left(\partial^2_R+\frac{1}{R}\partial_R+\frac{1}{R^2}\partial^2_\phi-\frac{Z}{R}+\frac{\alpha}{4R^2\cos^2(k\phi/2)}+
\frac{\beta}{4R^2\sin^2(k\phi/2)}\right)\Psi={\cal E}\Psi.
\end{gather}

Now note that if we divide both sides of expression (\ref{TTWham1}) by $r^2$, rearrange terms,
and  make the change of variables $R=r^2$, $\phi=2\theta$, then this expression is identical to  (\ref{deformedKepler})
with the identif\/ications
\begin{gather}\label{identifications}
\frac{\omega^2}{4}\leftrightarrow {\cal E}, \qquad \frac{E}{4}\leftrightarrow Z.\end{gather}
 Thus (\ref{deformedKepler}) is a St\"ackel
transform of (\ref{TTWham1}) and the two systems are St\"ackel equivalent, \cite{KMP10,PW2010, KKM10b}.

The principal observation that we need to make is that both systems have exactly the same formal eigenfunctions (\ref{sepeigenfunctions}),
modulo variable substitution and identif\/ications (\ref{identifications}), and exactly the same separation equations
and recurrence formulas. Thus, substituting into (\ref{energyev}) we see that the energy levels for the
deformed Kepler--Coulomb system are
\begin{gather*}
{\cal E}= \frac{Z^2}{\left(2(m+nk)+1+(a+b+1)k\right)^2}.
\end{gather*}
In the expressions for the raising and lowering operators of the TTW system we replace the operator $E\sim H$ by the constant $4Z$
 to get the the raising and lowering operators for the deformed  Kepler--Coulomb system. Similarly
the structure equations for the Kepler--Coulomb system are obtained by simple permutations
\[
\omega^2\leftrightarrow 4 H', \qquad H\leftrightarrow 4Z.
\]
We f\/ind
\begin{gather*}
 [L_2,L_4]=R,\\
 [L_2,R]=-8k^2q^2\{L_2,L_4\}-16k^4q^4L_4,\\
 [L_4,R]=8k^2q^2L_4^2-8k^2qP^{(-)}\big(16Z^2,L_2,4H',a,b\big),\\
\frac{3}{8k^2q^2}R^2+22k^2q^2L_4^2+\{L_2,L_4,L_4\}-4k^2qP^{(-)}\big(16Z^2,L_2,4H',a,b\big) \nonumber\\
\qquad{}  -12k^2P^{
(+)}\big(16Z^2,L_2,4H',a,b\big)=0. 
\end{gather*}

Similarly the operator $L_5$ can be added to the system.

\section{Discussion and conclusions}

We have developed a new method for verifying quantum superintegrability for 2D systems and given several applications to families of
such systems, most notably the caged anisotropic oscillator, the Tremblay--Turbiner--Winternitz system and the deformed Kepler--Coulomb
system and given new proofs of superintegrability for all rational $k$. The method relies on the assumption that the
system has a second order symmetry operator, so that the Schr\"odinger eigenvalue equation $H\Psi=E\Psi$  separates in a set of orthogonal
coordinates~$u_1$,~$u_2$ determined by the symmetry and so that the separated eigenfunctions $\Psi=U_1(u_1)U_2(u_2)$ satisfy
computable recurrence relations. In practice this means that
the separated solutions need to be of hypergeometric type. Then one  employs the recurrence relations to construct
operators that commute with $H$ on a~formal eigenbasis. We used our earlier developed canonical form for higher ($\ge 2$) order symmetry operators
to show that operators commuting with~$H$ on a~formal eigenbasis must  be actual symmetry operators, further that operator
identities verif\/ied on formal eigenbases must hold identically. Using this approach one can obtain explicit,
though complicated, expressions for the higher order symmetry operators.

We saw that in the case of the TTW potential our method didn't lead immediately to the lowest order generators,
 but that they could be found and expressed in terms of our raising and lowering operators. We have  no proof as yet that we have found the lowest order generator for all rational~$k$ but this is the case for all examples that we know. Provable determination of the maximal symmetry algebra is a topic for future research.
By acting on formal eigenbases we were able to compute  symmetry algebras for each of the systems we studied and show  that
these algebras were closed under commutation. One striking result of these computations was that the structure equations for the symmetries
were quite explicit and much simpler that the expressions for the symmetry operators themselves. In essence, one can determine the
structure equations without knowing the higher order generating symmetries!

In each case the action of the symmetries on a formal eigenbasis led us to a simple  model of the associated symmetry algebra and its
representations, in terms of dif\/ference operators in one variable. This greatly simplif\/ies the analysis of the structure equations,
classif\/ication of irreducible representations of the symmetry algebra and determination of the spectral properties of the generating
symmetries. Of course adding the generator $L_5$ to the algebra complicates these models, though by construction it can always be realized as a dif\/ference operator. This is an issue for future research. Such models have independent interest \cite{KMP2007, KMP2008, KMP2010}.

Finally, we gave an example of the use of the St\"ackel transform to map one superintegrable family of systems to another, while preserving
the structure equations. In this case we mapped the TTW system to a deformed Kepler--Coulomb system and determined,
for the f\/irst time,  the structure equation for the Kepler--Coulomb system.

It is interesting that this approach  to quantum superintegrability with structure results for symmetry algebras has preceded the classical approach; usually the reverse is true.  To our knowledge, the closure of  classical symmetry algebras has not been proven  for the systems considered here. We are actively investigating the classical analog of the quantum construction.
It is clear that these methods have much greater applicability than just the examples treated here. Indeed,  all of the
systems studied
in \cite{KKM10b} could be so analyzed. Further all our methods can clearly be extended to systems in dimensions $>2$. There appears to be no
obstacle other than growing complexity.

When it applies, the recurrence relation method is much simpler than our earlier introduced canonical form method for verif\/ication
of quantum superintegrability and it provides much more information, including the structure equations. However, the recurrence method
requires a~detailed knowledge of recurrence relations obeyed by the separated solutions and
there are examples of superintegrable systems
where no such relations appear to exist \cite[Section 4.2]{KMPog2006}. Thus the canonical form method for higher order
 symmetry operators
seems to be more general. Furthermore, we used the canonical form at a crucial point in the recurrence approach to show
that computations
valid on formal eigenbases actually held identically.
The approaches are mutually complementary.

\pdfbookmark[1]{References}{ref}
\LastPageEnding

\end{document}